\documentclass[a4paper,12pt]{report}
\usepackage{graphicx}

\begin{document}

\begin{center}
June 9, 2004\\
\begin{large}
{\bfseries Tunnelling through a semiconducting spacer: complex band predictions vs. thin film calculations}\\
\vspace{.4cm}
Jer\'onimo Peralta Ramos $^{a,*}$, Juli\'an Milano $^b$ and Ana Mar\'ia Llois $^a$\\
\end{large}
\vspace{0.2cm}
\begin{small}
{\itshape
$^a$ Departamento de F\'isica, Centro At\'omico Constituyentes, Comisi\'on Nacional de Energ\'ia At\'omica\\
$^b$ Centro At\'omico Bariloche, Comisi\'on Nacional de Energ\'ia At\'omica}\\
\end{small}
\end{center}
\vspace{1cm}

Using a simple tight-binding model, we compare the limitations of the tunnelling predictions coming out of the complex band structure of a semiconductor with the output of thin film calculations done for the same semiconducting spacer but considering it to be of finite width, and sandwiched by metallic electrodes. The comparison is made as a function of spacer width and interfacial roughness.
\\
KEY WORDS: tunnelling, complex bands, hybrid structures.
\\
\vspace{1cm}
\\
\begin{small}
$^{*}$ Corresponding author
\\ 
Centro At\'omico Constituyentes, Av. Gral. Paz 1650, San Mart\'in, Bs. As., Argentina
\\
TEL: (54-11) 6772-7007 FAX: (54-11) 6772-7121
\\
peralta@cnea.gov.ar
\end{small}
\newpage

A magnetic tunnel junction (MTJ) consists of a thin non-conducting spacer sandwiched by ferromagnetic electrodes. It is experimentally observed that the current through it depends on the relative alignment of the spins of the electrodes. This phenomenon is called tunnelling magnetoresistance (TMR) [1]. 
In the past few years MTJs have aroused considerable interest due to their potential applications in spin-electronic devices such as magnetic sensors and magnetic random-access memories. 
The diversity of the physical phenomena which govern the functioning of these devices makes MTJs also very attractive from a fundamental physical point of view [1,2].

The purpose of this work is to compare the information obtained from complex 
band calculations with what is obtained from thin film calculations, both for perfect and for roughed metal/spacer interfaces.

Mavropoulous {\it et al} showed that tunnelling through insulators and semiconductors can be understood qualitatively in terms of metal-induced gap states and that the framework to investigate them is the complex band structure of the insulator or semiconductor in the gap region, which is obtained from a bulk calculation independent of the characteristics of the electrodes [3].

For an infinite solid, boundary conditions impose a constraint on the wave vectors: they must be real in order to satisfy Schr$\ddot{o}$dinger's equation (with real energies). If the solid is periodic in one dimension but finite in the second one (for simplicity we assume a 2D solid, but the arguments hold for the 3D case as well), the wave vector parallel to the interface (let's say $k_y$) must still be real, but the component normal to the interface may be complex and still correspond to real energies ($k_x=q+i\kappa$, with $q$ and $\kappa$ real). The real energies corresponding to complex wave vectors are called {\bfseries complex bands}, and their eigenstates $\psi_{\vec{k}}\propto e^{-\kappa x}$ decay exponentially {\it into} the solid, $\kappa>0$ being the decay parameter. The solutions with $\kappa < 0$ represent eigenstates with ever growing amplitudes ($\propto e^{\left|\kappa\right| x}$), and must be discarded as solutions. 

Even though evanescent states can exist only near a surface or interface, their general properties can be derived from the bulk if one formally allows $k_x$ to become complex. The reason is that the interface-induced changes in the charge density and potential are confined to the first few monolayers, but the evanescent wave functions themselves extend over many layers into the crystal, thus representing solutions to the bulk Schr$\ddot{o}$dinger's equation [3].
 
Insulators or semiconductors can host evanescent states with their energy in the band gaps. It is known that if a metal and an insulator or semiconductor are put into contact the Fermi energy $E_F$ lies in the gap of the non-metal [4]. If we now have a trilayer of the type metal/insulator or semiconductor/metal, and the spacer thickness is small enough, electrons from one electrode are able to tunnel though the insulator, since metal eigenstates with energy close to $E_F$ can match evanescent states in the barrier which themselves match states on the other lead. In this context TMR is explained in terms of the different matching that occurs between metal and spacer states for the majority and minority spin channels [2,3,5,6].
 
Since the complex band structure of a spacer is independent of the leads attached to it, a series of questions arise: Are the tunnelling predictions extracted from complex band calculations reliable as compared to thin film calculations, which take into account the  interfaces and the finite size of the spacer? If this is so, does this correspondance break down, for example, for sufficiently thin spacers? 
 
To answer these questions, we consider a 2D semi-infinite model semiconductor whose structure is given by a square Bravais lattice of constant $a=3.2$ \AA ~with two atoms per unit cell, as shown in Fig. 1.
Each semiconductor layer consists of an infinite array of atoms A and B in the $y$ direction. The spacer is modelled by a tight-binding paramagnetic Hamiltonian with one $s$ orbital per site and hopping parameter $t_{AB}=0.2$ eV between first and second nearest-neighbours. The on site energy of atoms A is $E_A=1$ eV, and that corresponding to atoms B is $E_B=1.5$ eV. With these parameters, a band gap of $0.5$ eV is obtained. In what follows, we call `layer' to two successive monolayers, one with atoms of type A and the other with atoms of type B.

For each real value of $k_y$ and complex values of $k_x$ we diagonalize the bulk Hamiltonian that describes the spacer and find those $k_x$ values which yield real energies very close to a fixed energy in the middle of the gap of the semiconductor. This fixed energy is the Fermi energy of the metal leads, to be used later.

Fig. 2 shows the complex bands of our spacer for $q=\pi/a$ and $k_y=0$. In Fig. 2 we show only that complex band corresponding to the smallest $\kappa > 0$, which gives the maximum penetration into the spacer. 

Fig. 3 shows the value of the decay parameter $\kappa$ vs. $k_y$, for $q=\pi/a$. The decay parameter is equal to $(\pi/N a)$, where $N$ is the number of layers the evanescent state penetrates into the spacer. According to the complex band structure of the spacer, the maximum penetration is slightly larger than 5.5 layers and occurs for normal incidence ($k_y=0$). The penetration of the states into the barrier decreases with increasing $k_y$, and tends to zero for $k_y=\pi/a$.

In order to check the validity of this maximum penetration length, we consider the same semiconducting spacer, but now of finite width, sandwiched between metallic layers of the same structure. For simplicity we consider them to be paramagnetic, but the results are valid for magnetic ones as well. Epitaxial growth of the spacer on the metallic leads is assumed. The on site energy of the metallic atoms of type C and D is $E_C=2.8$ eV and $E_D=2.9$ eV respectively, while the corresponding hopping parameters between first and second nearest-neighbours is equal to $1$ eV (see Fig. 4). The Fermi energy $E_F$ is chosen to be $1.3$ eV and falls in the middle of the spacer's band gap.
  
To simulate interfacial roughness, the semiconducting atoms on the interfaces are replaced by atoms whose on site energy $E_i ~(left)$ and $E_i ~(right)$ is variable (Fig. 4). Taking $E_i ~(left)=E_A$ and $E_i~ (right)=E_B$ corresponds to perfect interfaces.

For a fixed number of metal layers we vary the number of spacer layers and for each value of $k_y$ we diagonalize the Hamiltonian obtaining the corresponding eigenstates and eigenenergies. Of these energies, we retain only those which are very close to the Fermi energy of the leads, and plot the amplitude of these states on the atomic sites along the $x$ direction. This allows us to investigate the penetration of these states into the spacer. The results are practically independent of the number of metal layers put at each side of the spacer, if the number of metal layers is larger than 10, so we fix it equal to 30, which is large enough for the leads to be considered semi-infinite.

Fig. 5 shows an example of a state decaying into the barrier in the case of a spacer of 10 layers.
The main result of this study is that for a number of spacer layers larger than the maximum penetration infered from the complex bands ($\sim$ 5.5), the dependence on $k_y$ of the number of spacer layers penetrated by the decaying states is in perfect agreement with the predictions of the complex band structure.
 
The situation is different when the number of spacer layers is smaller than 5.5. In this case the dependence on $k_y$ of the number of penetrated spacer layers is not the one given by the complex bands. The penetration is complete for values of $k_y$ in the range $[0,0.9] ~\pi/a$, in contradiction with the complex bands predictions. Fig. 6 shows an example of a state with $k_y=0.86 ~\pi/a$ penetrating completely through 3 spacer layers. According to the complex band structure, this state should penetrate only 1.5 layers (see Fig. 3).

Varying the on site energies of the impurities on the left and right interfaces, we find that the penetration of states into the spacer is maximum for perfect interfaces ($E_i~ (left) = E_A$ and $E_i~ (right) = E_B$). Fig. 7 shows the result of leaving $E_i ~(right)$ constant and equal to $E_B$ and taking $E_i~ (left) = E_A+\eta$ with $\eta$ in the range $[-5,5]$ eV. Fig. 7 corresponds to 10 spacer layers. The penetration obtained when the interfaces are not perfect is smaller than the one predicted by the complex bands of the spacer. This result indicates that the complex band structure of the spacer cannot account for interface roughness effects, not even in the case of thick spacers.

In summary, by studying a simple system we have found that the knowledge of the complex band structure of the spacer is not enough to fully understand and predict the tunnelling properties of MTJs, especially for thin spacers. The limit between thin and thick spacers is set by the maximum penetration given by the complex bands of the spacer. The agreement between the predictions stemming out of the complex bands and those coming from thin film calculations is remarkably good for thick spacers and very poor for thin ones, in the case of perfect interfaces.

In the case of rough interfaces, the penetration is always smaller than the one predicted by the complex bands. 

In agreement with the results of recent theoretical studies [7,8], we conclude that it is critical to consider the metal/spacer interfaces, not only for thin but also for thick spacers, and that the tunnelling predictions coming from the complex band structure of the spacer are reliable only for the case of perfect metal/spacer interfaces. Therefore, TMR values obtained from complex band calculations are to be considered with care.

This work was partially funded by UBACyT-X115, Fundaci\'on Antorchas and PICT 03-10698. Ana Mar\'ia Llois belongs to CONICET (Argentina). 
\newpage

\begin{center}
References\\
\end{center}
$[1]$ X-G Zhang and W. H. Butler, J. Phys.: Condens. Matter {\bfseries 15} (2003) R1603-R1639\\
$[2]$ Evgeny Y. Tsymbal, Oleg N. Mryasov and Patrick R. LeClair, J. Phys.: Condens. Matter {\bfseries 15} (2003) R109-R142\\
$[3]$ Ph. Mavropoulos, N. Papanikolaou and P. H. Dederichs, Phys. Rev. Lett. {\bfseries 5} (2000) 1088\\
$[4]$ V. Heine, Phys. Rev. {\bfseries 138} (1965) A1689\\
$[5]$ W. H. Butler, X-G. Zhang, T. C. Schulthess and J. M. MacLaren, Phys. Rev. B {\bfseries 63} (2001) 054416\\ 
$[6]$ J. M. MacLaren, X.-G. Zhang, W. H. Butler and Xindong Wang, Phys. Rev. B {\bfseries 59} (1999) 5470\\
$[7]$ O. Wunnicke, N. Papanikolaou, R. Zeller, P. H. Dederichs, V. Drchal and J. Kudrnovsky, Phys. Rev. B {\bfseries 65} (2002) 064425\\
$[8]$ K. D. Belashchenko, E. Y. Tsymbal, M. van Schilfgaarde, D. A. Stewart, I. I. Oleynik and S. S. Jaswal, Phys. Rev. B {\bfseries 69} (2004) 174408\\

\newpage

Figure 1\\
Semi-infinite spacer with a square Bravais lattice and two atoms per unit cell, A and B. The system is periodic in the $y$ direction.
\\

Figure 2\\
Complex band structure of the semi-infinite spacer, for $q=\pi/a$ and $k_y=0$.
\\

Figure 3\\
Imaginary part of $k_x=q+i\kappa$ as a function of $k_y$, for $q = \pi/a$. The decay parameter $\kappa$ is equal to $(\pi/N a)$, where $N$ is the number of layers the evanescent state penetrates into the spacer. The maximum penetration, 5.5 layers, occurs for $k_y=0$, and the penetration decreses with increasing $k_y$, going to zero as $k_y$ approaches $\pi/a$.
\\

Figure 4\\
Semiconducting spacer of finite width sandwiched between metallic layers. The figure corresponds to 3 spacer layers and 2 metal layers. The on site energies $E_i~ (left)$ and $E_i ~(right)$ are variable. Taking $E_i ~(left)=E_A$ and $E_i ~(right)=E_B$ corresponds to perfect interfaces.
\\

Figure 5\\
Decaying state amplitude as a function of position for 10 spacer layers, with $k_y=0.4 ~\pi/a$. The penetration is $\sim 5$ layers, in agreement with the predictions of the complex band structure of the semiconductor.
\\

Figure 6\\
Decaying state amplitude as a function of position for three spacer layers, with $k_y=0.86~ \pi/a$. The penetration is complete, in disagreement with the predictions of the complex band structure of the semiconductor.
\\

Figure 7\\
$E_i ~(right)$ is constant and equal to $E_B$ and $E_i~ (left) = E_A+\eta$, with $\eta$ in the range $[-5,5]$ eV. The penetration obtained when the interfaces are not perfect is smaller than the one predicted by the complex bands of the spacer.
\\
\newpage

\begin{figure}
\scalebox{1}{
\includegraphics{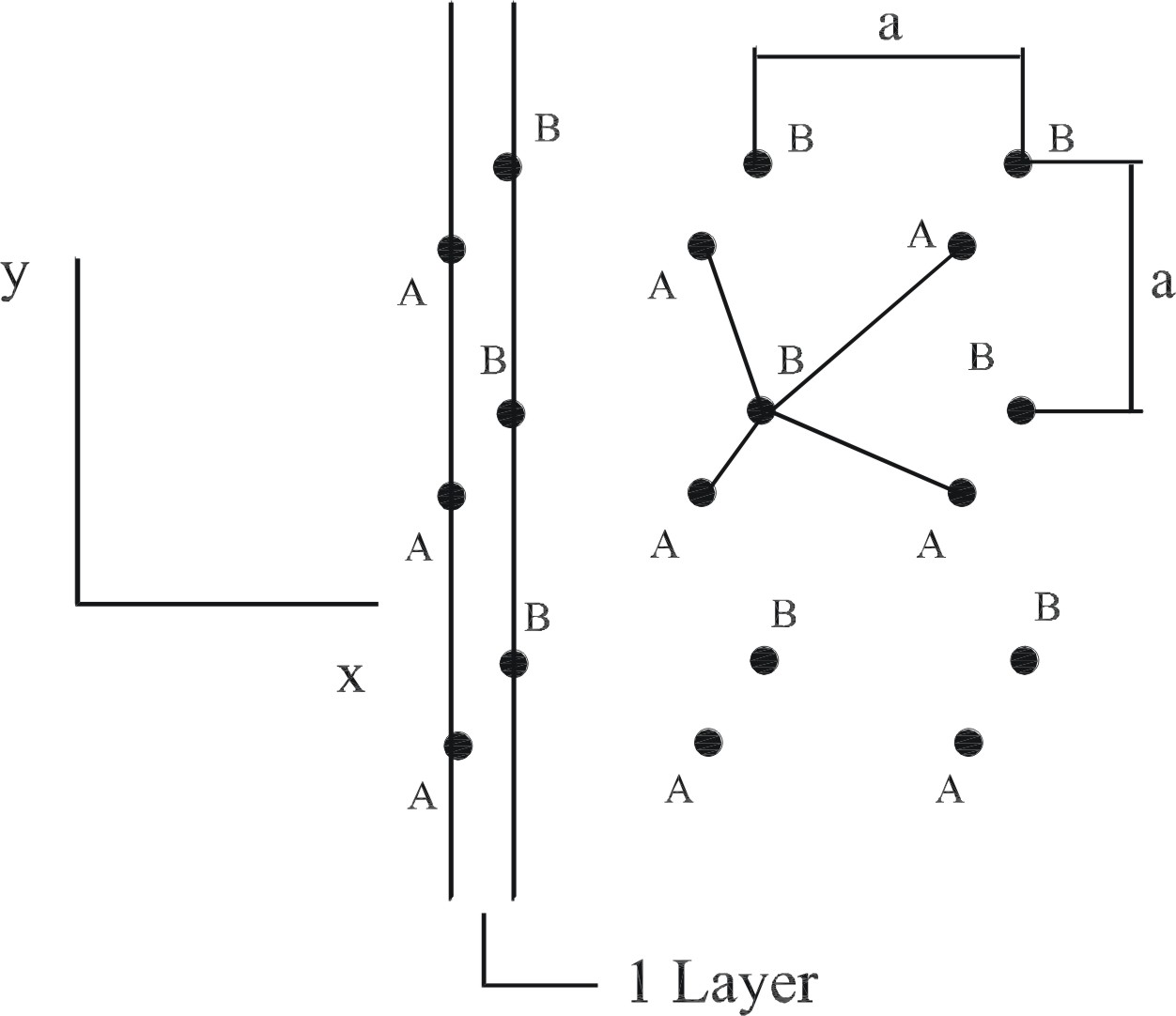}}
\caption{}
\end{figure}

\begin{figure}
\scalebox{0.85}{
\includegraphics{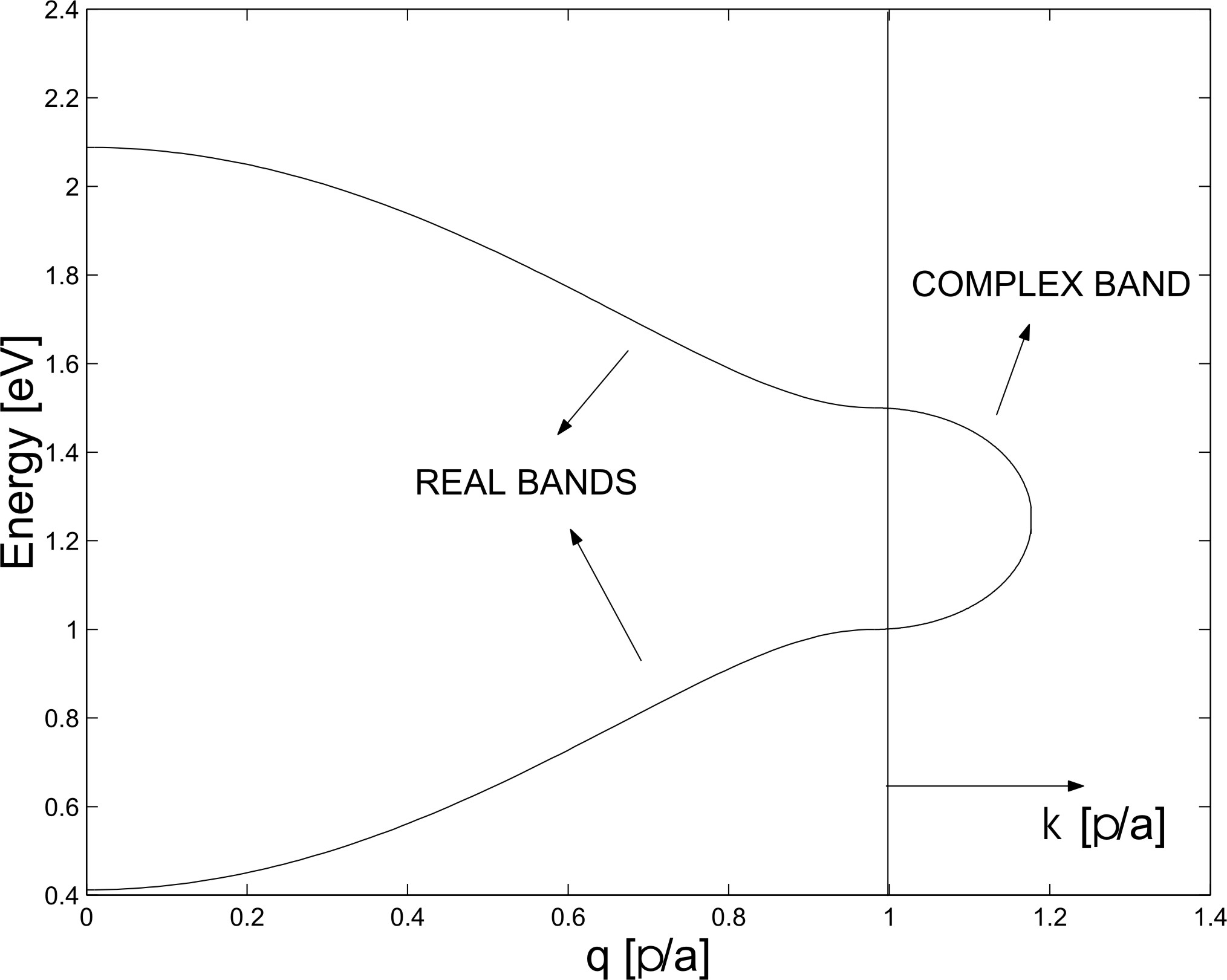}}
\caption{}
\end{figure} 

\begin{figure}
\scalebox{0.85}{
\includegraphics{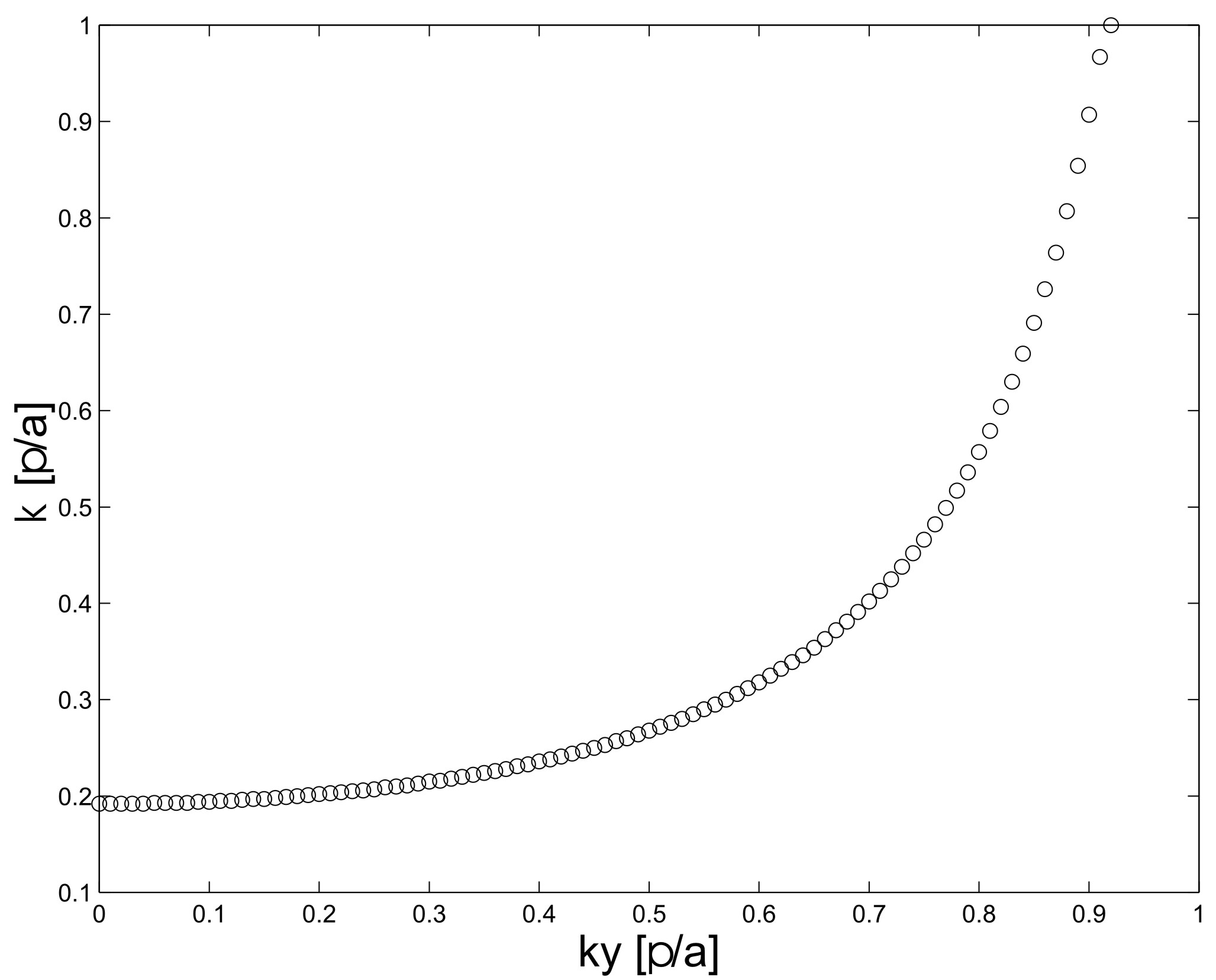}}
\caption{}
\end{figure} 

\begin{figure}
\scalebox{0.7}{
\includegraphics{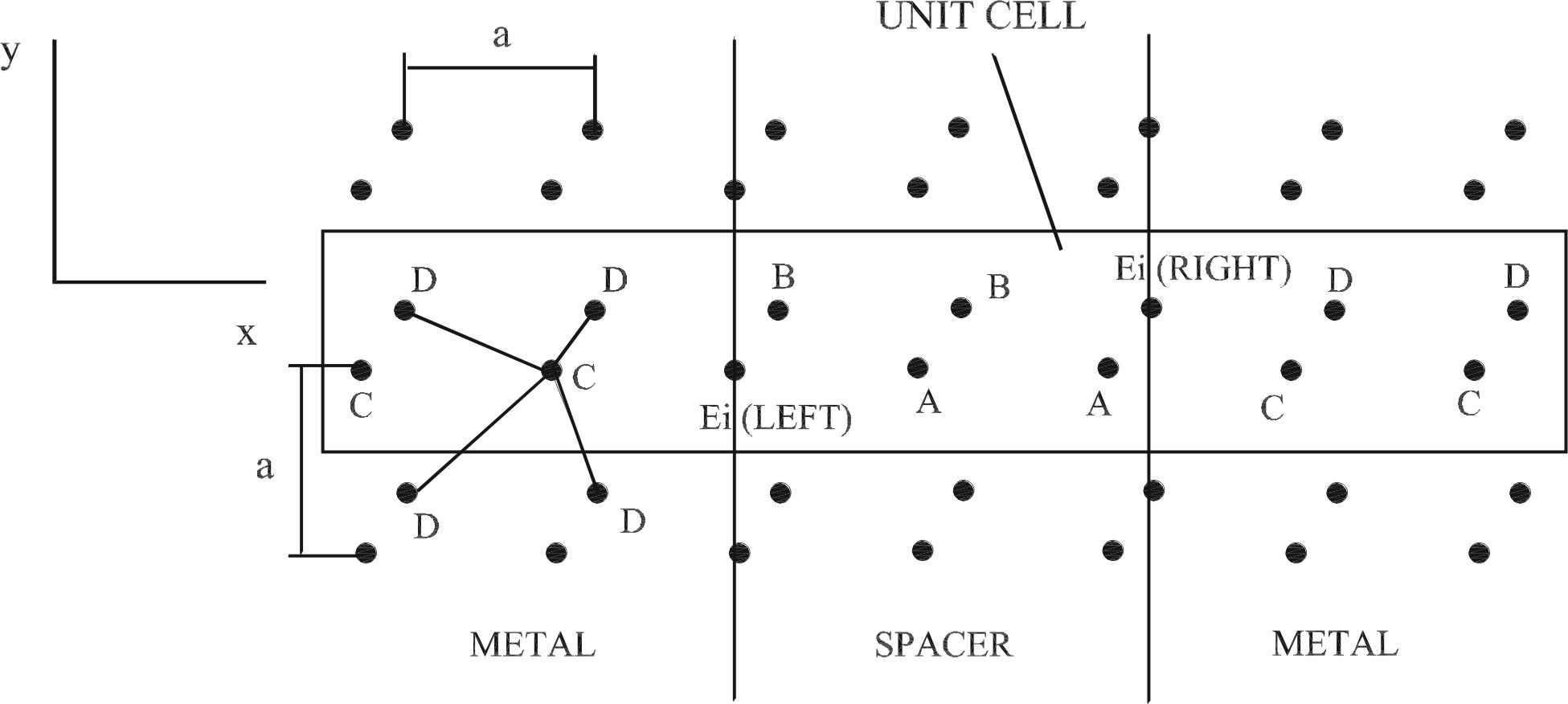}}
\caption{}
\end{figure} 

\begin{figure}
\scalebox{0.85}{
\includegraphics{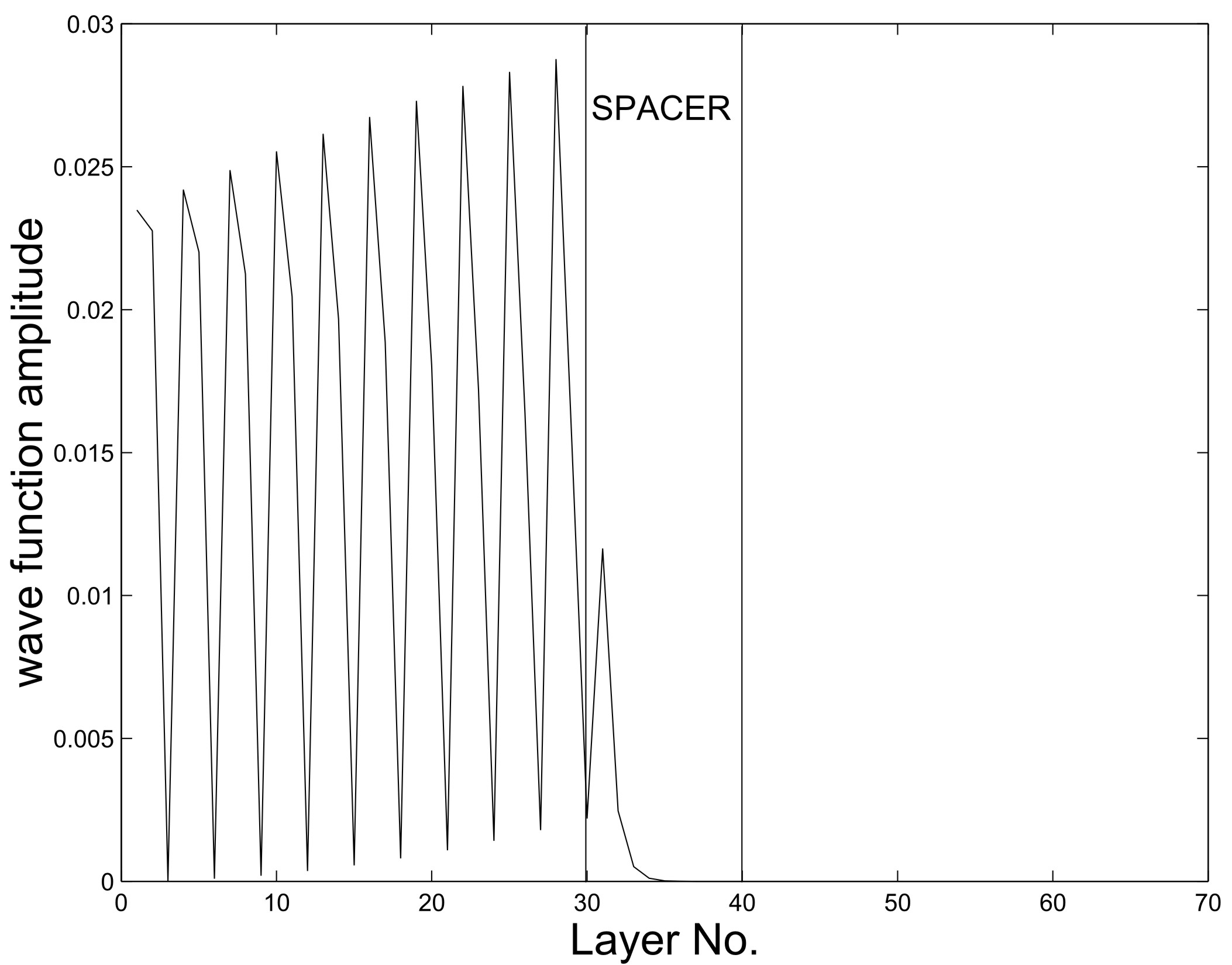}}
\caption{}
\end{figure} 

\begin{figure}
\scalebox{0.85}{
\includegraphics{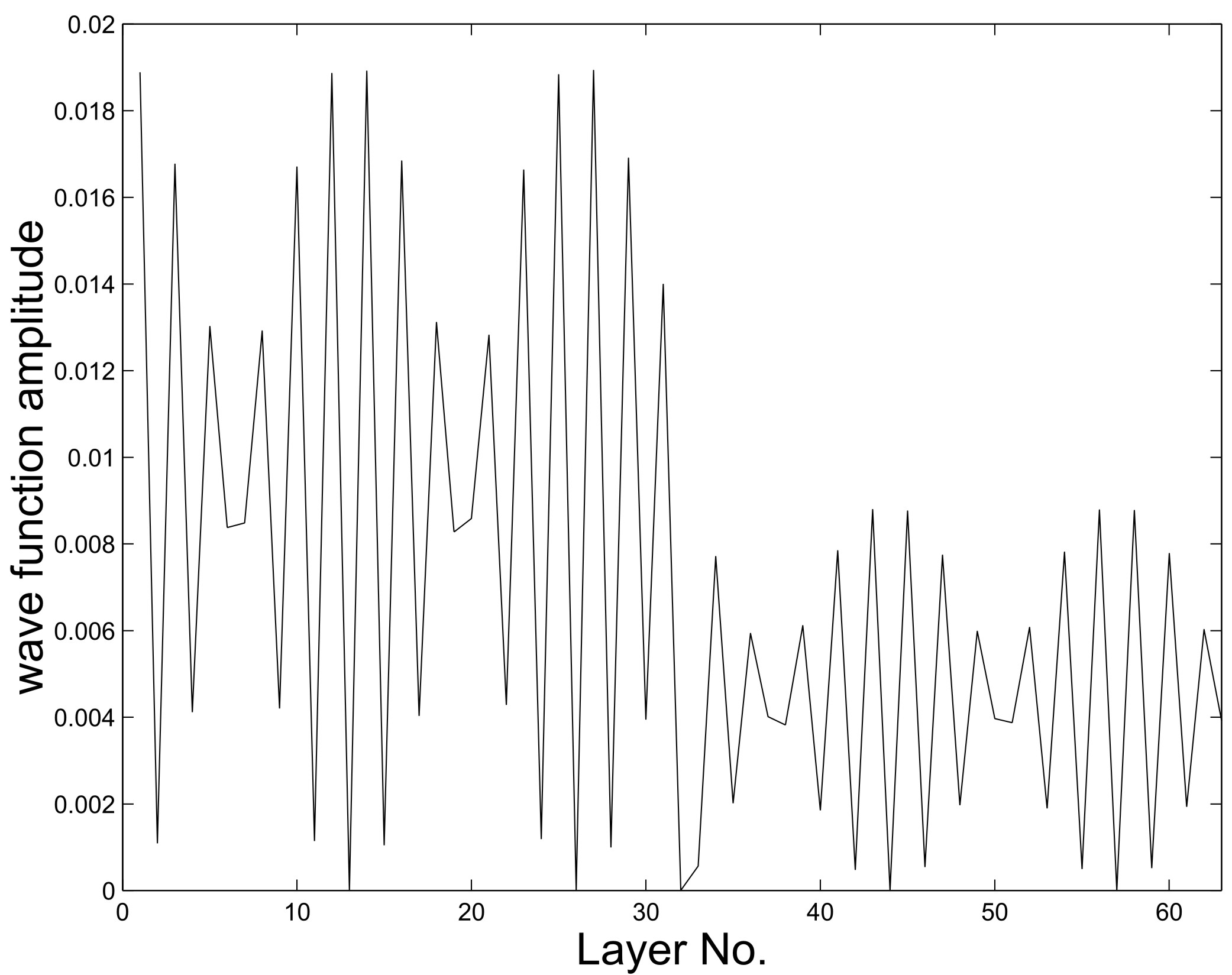}}
\caption{}
\end{figure} 

\begin{figure}
\scalebox{1.2}{
\includegraphics{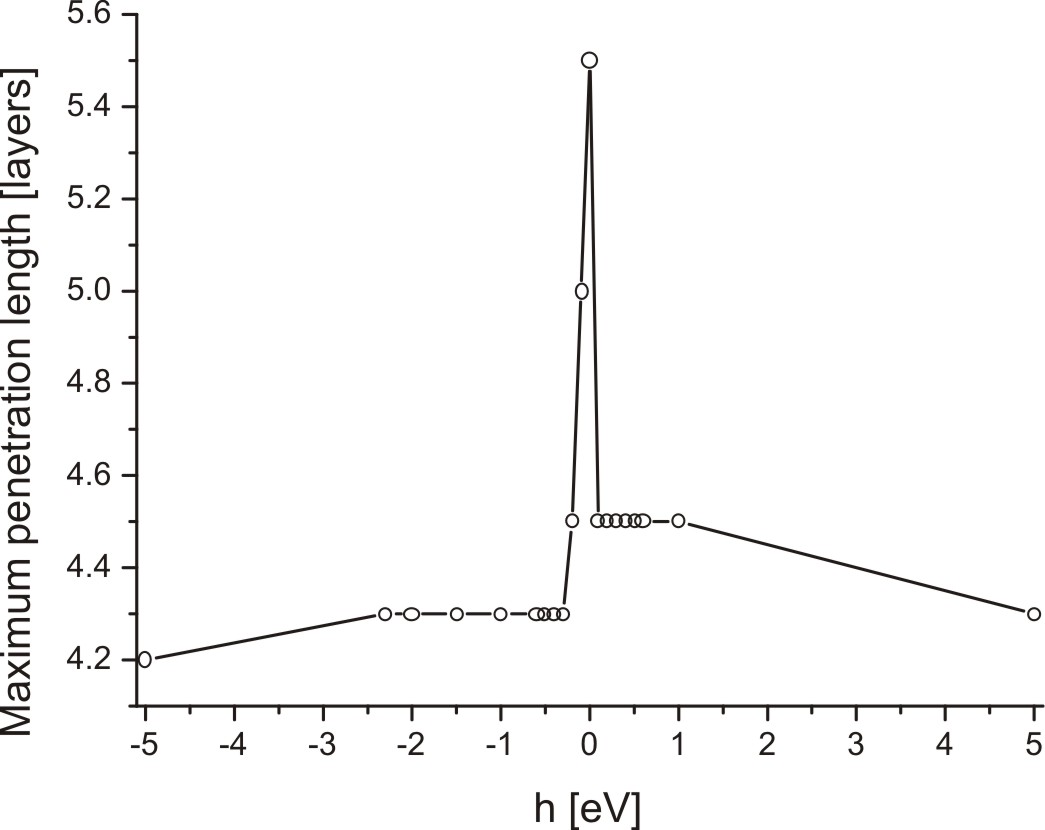}}
\caption{}
\end{figure}

\end{document}